\documentclass[epj]{webofc}
\usepackage[varg]{txfonts}   
\wocname{EPJ Web of Conferences}
\woctitle{CONF12}
\begin{document}
\selectlanguage{english}
\title{Domain wall network as QCD vacuum: confinement, 

chiral symmetry, hadronization}
%
%

\author{Sergei N. Nedelko\inst{1}\fnsep\thanks{\email{nedelko@theor.jinr.ru}} \and
        Vladimir E. Voronin\inst{1} 
}

\institute{Bogoliubov Laboratory of Theoretical Physics, JINR, Dubna, Russia
}

\abstract{%
An approach to QCD vacuum as a medium describable in terms of statistical ensemble of almost everywhere
 homogeneous Abelian (anti-)self-dual gluon fields is  reviewed. These fields play the role of the
  confining medium for color charged fields as well as underline the mechanism of realization of chiral
     $SU_{\mathrm L}(N_f)\times SU_{\mathrm R}(N_f)$ and $U_{\mathrm A}(1)$ symmetries. Hadronization
    formalism based on this ensemble leads to manifestly defined quantum effective meson action. Strong,
     electromagnetic and weak interactions of mesons are represented in the action in terms of nonlocal 
     $n$-point interaction vertices given by the quark-gluon loops averaged over the background ensemble.
      Systematic results for the mass spectrum and decay constants of radially excited light, heavy-light
       mesons and heavy quarkonia  are presented. Relationship of
        this approach to the  results of functional renormalization group and Dyson-Schwinger equations, and
         the picture of harmonic confinement is briefly outlined.
}
\maketitle
\section{Introduction}
\label{intro}

The idea of four-dimensional harmonic oscillator as a tool for universal description of  the Regge spectrum of hadron masses was formulated long time ago by Feynman,  Kislinger, and Ravndal~\cite{Feynman}  and later  re-entered  the  discussion about quark confinement  several times  in  various ways. In particular,  Leutwyler and Stern  developed  the formalism devoted to the  covariant description of bilocal meson-like fields  combined with the idea of harmonic confinement~\cite{Leutwyler:1977vz,Leutwyler:1977vy, Leutwyler:1977pv, Leutwyler:1977cs, Leutwyler:1978uk}.    In recent years,  the approaches to confinement based on the soft wall AdS/QCD model with harmonic potential demonstrated an impressive phenomenological success.

The approach to QCD vacuum and hadronization presented here  has been developed in a series of papers~\cite{EN, EN1,NK1,NK4,NK6,NV,NV1}. It   clearly 
incorporates the idea of harmonic confinement both in terms of elementary color charged fields and the composite colorless  hadron fields. The distinctive feature of the approach is that it links the concept of harmonic confinement  and Regge character of hadron mass spectrum to the specific class of nonperturbative gluon configurations -- almost everywhere homogeneous Abelian (anti-)self-dual gluon fields. 
A  close interrelation of the   Abelian (anti-)self-dual fields and the hadronization based on harmonic confinement   can be read off the  papers~\cite{Pagels, Minkowski, Leutwyler1, Leutwyler2,  Leutwyler:1977vz,Leutwyler:1977vy, Leutwyler:1977pv, Leutwyler:1977cs, Leutwyler:1978uk}.
In brief,  the line of arguments is as follows (for more detailed exposition see~\cite{NV,NV1}). 

An important starting point is due to the  observation of  Pagels and Tomboulis~\cite{Pagels} that Abelian self-dual fields describe a  medium that is  infinitely stiff to small gauge field fluctuations, i.e. the  wave solutions for the effective quantum equations of motion are absent. This feature was interpreted as suggestive of confinement of color. 
Strong argumentation in favour of the Abelian (anti-)self-dual homogeneous field as a candidate for the global nontrivial minimum of the effective action originates from the papers \cite{Minkowski,Leutwyler2,NG2011,Pawlowski,George:2012sb}.   Leutwyler has shown  that the constant gauge field is stable   against small quantum fluctuations only if it is Abelian (anti-)self-dual covariantly constant field  \cite{Minkowski,Leutwyler2}. Nonperturbative calculation of the effective potential within the functional renormalization group \cite{Pawlowski} supported the earlier one-loop results on existence of the nontrivial minimum of the effective action for the Abelian (anti-)self-dual field.

\begin{figure}
\centering
\sidecaption
\includegraphics[width=0.35\textwidth]{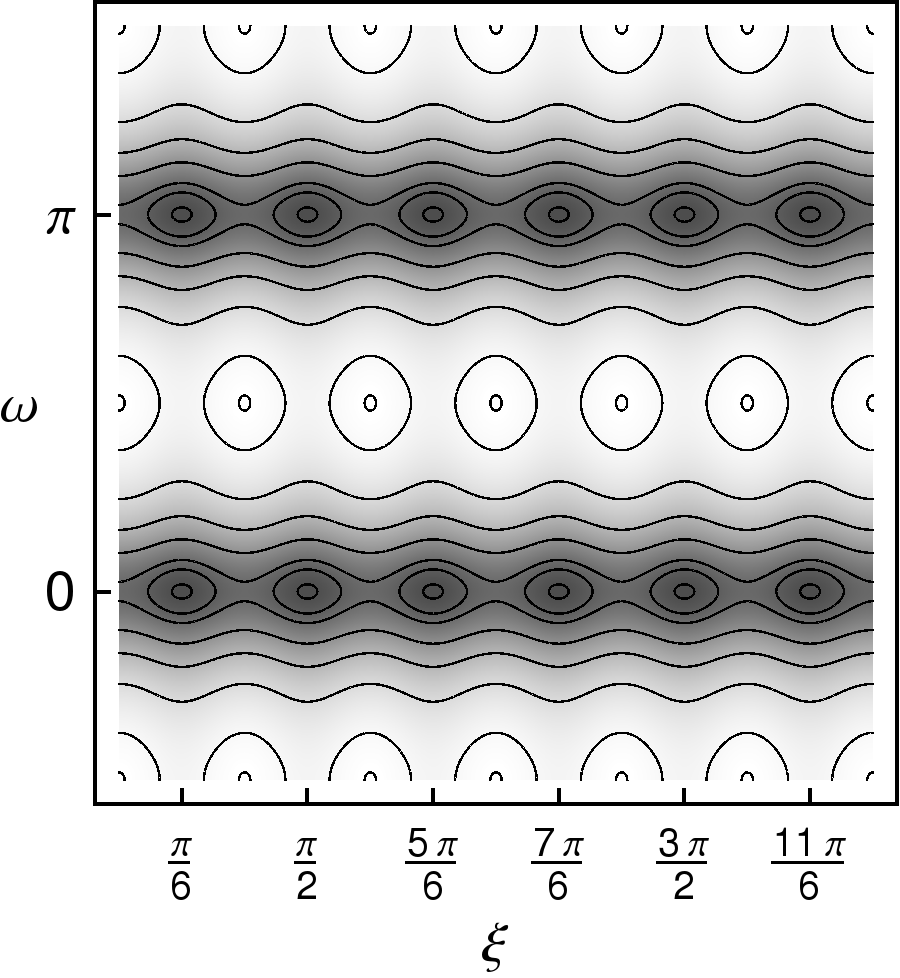}
\caption{Effective potential  as a function of the angle $\omega$ between chromomagnetic and chromoelectric field  and the mixing angle $\xi$  in the Cartan subalgebra. The minima in the dark gray regions correspond to the Abelian  (anti-)self-dual  configurations and form a periodic  structure labelled  by integer indices $(kl)$  in Eq.~\eqref{minima} (for more details see  \cite{NK1,NG2011,NV}).\label{effpotxiomega}}
\end{figure}
\begin{figure}[h!]
\centering
\sidecaption
\includegraphics[width=.2\textwidth]{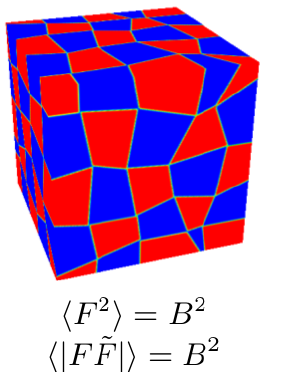}\includegraphics[width=.175\textwidth]{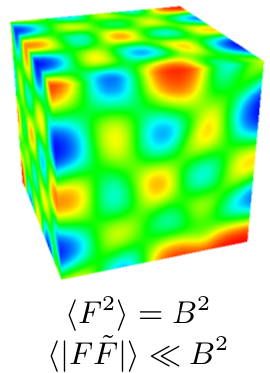}
\caption{Topological charge density for domain wall networks with different values of the wall width. The left picture is an example of confining almost everywhere homogeneous Abelian (anti-)self-dual fields. Red (blue) color corresponds to the self-dual field (anti-self-dual), green -- pure chromomagnetic field. The right plot represents the case of preferably pure chromomagnetic field when the topological charge density is nearly zero and color charged quasiparticles can be excited thus indicating deconfinement (for more details see \cite{NV}).\label{cubes}}
\end{figure}

The eigenvalues of the Dirac and Klein-Gordon operators in the presence of Abelian self-dual field are purely discrete, and the corresponding eigenfunctions  of quarks and gluons charged with respect to the  background  are of the bound state type. This is a consequence of the 
fact that these operators contain the four-dimensional harmonic oscillator.   Eigenmodes  of the color charged fields have no (quasi-)particle interpretation but describe field fluctuations decaying in space and time. The consequence of this property is  that the momentum representation of the translation invariant part of the propagator of the color charged field in the background of homogeneous  (anti-)self-dual Abelian gauge field is an entire analytical function. 
The absence of pole in the propagator was treated as the absence of the particle interpretation of the charged field~\cite{Leutwyler1}. 
 However,  neither the homogeneous Abelian (anti-)self-dual
field itself nor  the form of gluon propagator in the presence of this background
had the clue to linear quark-antiquark potential.  Nevertheless, the analytic structure of the gluon and quark propagators and assumption about the randomness of the background field ensemble led both to the area law for static quarks and the Regge spectrum for light hadrons.

The model of hadronization developed in~\cite{EN1,EN, NV,NV1} indicated that the spectrum of mesons displayed the Regge character  both with respect to total angular momentum  and radial quantum number of the meson.  The  reason for confinement of a single quark and Regge spectrum of mesons turned out to be the same -- the analytic properties of quark and gluon propagators.  In this formalism any meson looks much more like  a complicated collective excitation of a medium (QCD vacuum) involving quark, antiquark and gluon fields than a nonrelativistic quantum mechanical bound state of charged  particles (quark and anti-quark) due to some potential interaction between them. Within this relativistic quantum field description the Regge spectrum of color neutral collective modes appeared as a "medium effect" as well as the suppression (confinement) of a color charged elementary modes.  

These observations have almost completed the quark confinement picture based on the random almost everywhere homogeneous Abelian (anti-)self-dual fields. Self-duality of the fields plays the crucial role in this picture. This random field ensemble represents a medium where the color charged elementary excitations exist as quickly decaying in space and time field fluctuations but the collective colorless excitations (mesons) can propagate as plain waves (particles).  Besides this dynamical color charge confinement,  a correct complete picture must include the limit of  static quark-antiquark pair with the area law for the temporal Wilson loop. In order to explore this aspect an  explicit construction of the random domain ensemble was suggested in paper~\cite{NK1}, and the area law for the Wilson loop was demonstrated by the explicit calculation. Randomness of the ensemble (in line with \cite{Olesen7}) and (anti-)self-duality of the fields are crucial for this result.

 The  character of meson wave functions   in hadronization approach~\cite{EN1} is fixed by the form of the gluon propagator in the  background  of  specific  vacuum gluon configurations.     These wave functions  are  very similar to the wave functions of the soft wall AdS/QCD with  quadratic dilaton profile and Leutwyler-Stern formalism. In all three cases one deals with the generalized Laguerre polynomials as characteristic for the radial meson excitations. Another interesting observation is that the form of Euclidean gluon and quark propagators in the presence of the Abelian (anti-)self-dual background  are in qualitative agreement with the  decoupling scenario of the infra-red behaviour of the propagators  in Dyson-Schwinger equations (DSE) and functional renormalization group (FRG) approaches and  lattice QCD results for the Landau gauge propagators.

The next section  is devoted to motivation of the approach based on the domain wall network gluon configurations.  xt of the spontaneous chiral symmetry breaking by the background field and four-fermion interaction.  
The structure of the effective meson action  and the results  for the  masses, transition and decay constants of various mesons   are presented in  section \ref{section2}. 
 In the  last section  we outline the relation of  gluon propagator in the model under consideration to  the results of FRG and DSE.

\begin{figure}[h!]
\centering
\includegraphics[scale=.8]{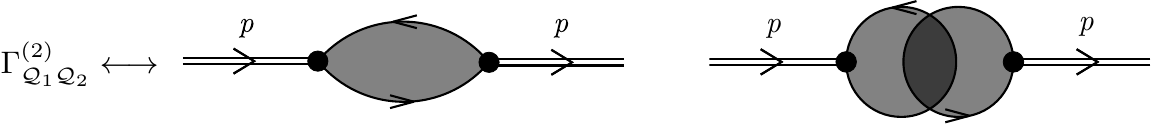}\\
\includegraphics[scale=.8]{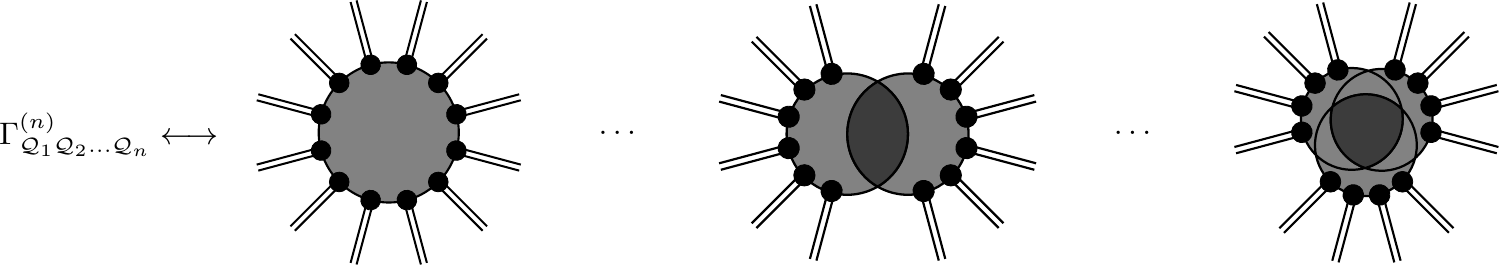}
\caption{Diagrammatic representation of nonlocal meson vertex functions. Light grey denotes averaging over background field, dark grey denotes correlation of loop diagrams by background field.\label{diagrams}}
\end{figure}

\begin{figure}[h]
\centering
\sidecaption
\includegraphics[width=9cm,clip]{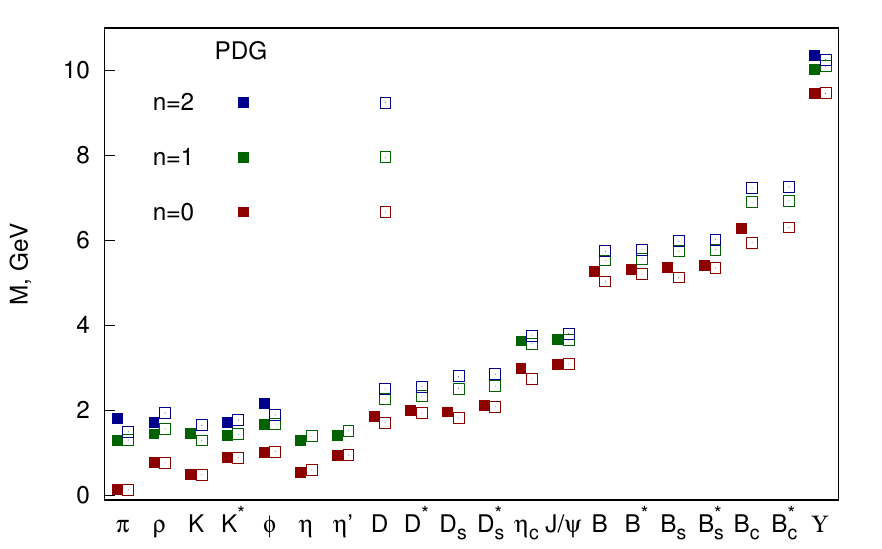}
\caption{The masses of various radially excited mesons. The same  values of parameters  were used for all
mesons shown in the figure. }
\label{fig-masses}       
\end{figure}

\section{Scalar gluon condensate and the effective action of QCD}
\label{section1}

The  phenomenological basis of the present approach is the assumption about existence of nonzero gluon condensates in QCD, first of all -- the scalar  condensate $\langle g^2F^2\rangle$.  In order to incorporate this condensate into the  functional integral approach to quantization of QCD one has to choose appropriate  conditions for  the functional space of gluon fields $A_\mu^a$ to be integrated over (see, e.g., Ref.\cite{faddeev}). 
Besides the formal mathematical  content, these conditions play the role of substantial physical input  which, together with the classical action of QCD, complements the statement of the quantization problem. In other words, starting with  the very basic  representation of the Euclidean  functional integral for QCD, 
one has to specify integration spaces $\mathcal{F}_A$ for gluon and  $\Psi$ for quark fields. 
Bearing in mind a nontrivial QCD vacuum structure encoded in various condensates, one have to define  $\mathcal{F}_A$  permitting  gluon fields with nonzero classical action density.
The gauge fields $A$ that satisfy this condition  have a potential to provide the vacuum with the whole variety of  condensates.

An analytical approach to definition and calculation of the functional integral can be based on separation of gluon modes  $B_\mu^a$ responsible for nonzero  condensates  from the small perturbations $Q_\mu^a$. This separation must be supplemented with gauge fixing.  Background gauge fixing condition $D(B)Q=0$ is the most natural choice. To perform separation, one inserts identity
\begin{equation*}
1=\int\limits_{{\cal B}}DB \Phi[A,B]\int\limits_{{\cal Q}} DQ\int\limits_{\Omega}D\omega \delta[A^\omega-Q^\omega-B^\omega]
 \delta[D(B^\omega)Q^\omega]
\end{equation*}
in the functional integral and arrives at 
\begin{eqnarray*}
Z &=&N'\int\limits_{{\cal B}}DB \int\limits_{\Psi} D\psi D\bar\psi\int\limits_{{\cal Q}} DQ \det[\mathcal{D}(B)\mathcal{D}(B+Q)]
\delta[\mathcal{D}(B)Q]e^{-S_{\rm QCD}[B+Q,\psi, \bar\psi]}\\
&=&\int\limits_\mathcal{B}DB \exp\{-S_\mathrm{eff}[B]\}.
\end{eqnarray*}

\begin{table}[htbp]
\centering
\caption{Model parameters fitted to the masses of $\pi,\rho,K,K^*, \eta', J/\psi,\Upsilon$ and used in  calculation of all other meson masses, decay and transition constants.}
{\begin{tabular}{@{}ccccccc@{}}
 \hline
$m_{u/d}$, MeV&$m_s$, MeV&$m_c$, MeV&$m_b$, MeV&$\Lambda$, MeV&$\alpha_s$&$R$, fm\\
\hline
$145$&$376$&$1566$&$4879$&$416$&$3.45$&$1.12$\\
\hline
\end{tabular}}
\label{values_of_parameters}
\end{table}

Thus defined quantum effective action $S_\text{eff}[B]$ has a physical meaning of the free energy of the quantum field system in the presence of the background  gluon field $B_\mu ^a$.  In the limit $V\to \infty$  global minima of  $S_\text{eff}[B]$ determine the class of  gauge field configurations  representing  the equilibrium state (vacuum) of the system. Quite reliable argumentation in favour of (almost everywhere) homogeneous Abelian  (anti-)self-dual fields as dominating vacuum configurations in pure gluodynamics was put forward by many authors~\cite{Pagels,Leutwyler2}.  
As it has already been mentioned in Introduction, nonperturbative calculation of QCD quantum effective action  within the functional renormalization group approach \cite{Pawlowski} supported the  one-loop result \cite{Pagels,Minkowski,Leutwyler2} and indicated the existence of a minimum of the effective potential for nonzero value of Abelian (anti-)self-dual homogeneous gluon field.

\begin{table}[h]
\centering
\caption{Decay and transition constants of various  pseudoscalar and vector mesons calculated through the diagrams shown  in Figs.~\ref{weak_decay_diagrams} and \ref{g_rho_gamma_diagrams}. }
{\begin{tabular}{@{}cccc|cccc@{}} \hline
Meson&$n$&$f_P^{\rm exp}$&$f_P$&Meson&$n$&$g_{V\gamma}$ 
\cite{PDG}&$g_{V\gamma}$\\
&&(MeV)& (MeV)&&&\\
\hline
$\pi$      &0 &130 \cite{PDG}        &140 & $\rho$&0&0.2&0.2 \\
$\pi(1300)$&1 &                      &29 & $\rho$&1&&0.053 \\
\hline
$K$        &0 &156 \cite{PDG}        &175  & $\omega$&0&0.059&0.067\\
$K(1460)$  &1 &                      &27   & $\omega$&1&&0.018\\
\hline
$D$        &0 &205 \cite{PDG}        &212  & $\phi$&0&0.074&0.071\\
$D$        &1 &                      &51   & $\phi$&1&&0.02\\
\hline
$D_s$      &0 &258 \cite{PDG}        &274  & $J/\psi$&0&0.09&0.06\\
$D_s$      &1 &                      &57  & $J/\psi$&1&&0.015\\
\hline
$B$        &0 &191 \cite{PDG}        &187  & $\Upsilon$&0&0.025&0.014\\
$B$        &1 &                      &55   & $\Upsilon$&1&&0.0019\\
\hline
$B_s$      &0 &253 \cite{Chiu:2007bc}&248  &  & &\\
$B_s$      &1 &                      &68   &  & &\\
\hline
$B_c$      &0 &489 \cite{Chiu:2007bc}&434  &  & &\\
$B_c$      &1 &                  &135  & &&\\
\hline
\end{tabular}
\label{constants}}
\end{table}

Ginzburg-Landau (GL) approach to the quantum effective action indicated a possibility of the domain wall network formation in  QCD vacuum resulting in the  dominating vacuum gluon configuration seen as an ensemble of densely  packed lumps of covariantly constant Abelian (anti-)self-dual field \cite{NK1,NG2011,NV,George:2012sb}. Nonzero scalar gluon condensate $\langle g^2F^a_{\mu\nu}F^a_{\mu\nu}\rangle$ 
postulated by the effective potential
leads 
to the existence  of twelve discrete degenerate global minima of the effective action (see Fig.\ref{effpotxiomega}), 
\begin{eqnarray}
&&\breve A_{\mu}\in\left\{\breve B^{(kl)}_{\mu}| \  k=0,1,\dots,5; \  l=0,1\right\}, \ \
\breve B^{(kl)}_{\mu}  = -\frac{1}{2}\breve n_k B^{(l)}_{\mu\nu}x_\nu, 
\nonumber\\
 && \tilde B^{(l)}_{\mu\nu}=\frac{1}{2}\varepsilon_{\mu\nu\alpha\beta} B^{(l)}_{\alpha\beta}=(-1)^l B^{(l)}_{\mu\nu},
\ 
\breve n_k = T^3\ \cos\left(\xi_k\right) + T^8\ \sin\left(\xi_k\right),
\ \
\xi_k=\frac{2k+1}{6}\pi,
\label{minima}
\end{eqnarray}
where $l=0$ and $l=1$ correspond to the self-dual and anti-self-dual field respectively,  matrix $\breve{n}_k$ belongs to Cartan subalgebra of $su(3)$ with six values of the angle $\xi_k$ corresponding to the boundaries of the Weyl chambers in the root space of $su(3)$. 
The minima are connected by the parity and Weyl group reflections.  
Their existence indicates that the system is prone to the domain wall formation.  To demonstrate the simplest example of domain wall interpolating between the self-dual and anti-self-dual Abelian configurations, one allows  the angle  $\omega$ between chromomagnetic and chromoelectric fields to vary from point to point in $R^4$  and restricts other degrees of freedom of gluon field to their vacuum values.
In this case Ginsburg-Landau Lagrangian  leads  to the  sine-Gordon equation for  $\omega$ with the standard
 kink solution (for details see Ref.~\cite{NG2011,NV}).
Away from the kink location vacuum  field is almost self-dual ($\omega=0$) or anti-self-dual ($\omega=\pi$).  Exactly at the wall it becomes purely chromomagnetic ($\omega=\pi/2$).  Domain wall network is constructed  by means of the kink superposition. 
Topological charge density distribution for a network of domain walls with different width is illustrated in Fig.\ref{cubes}.

Based on this construction, the measure of integration over the background field $B_\mu^a$ can be constructively represented as the infinite dimensional (in the infinite volume) integral over the parameters  of $N\to\infty$ domain walls in the network:  their positions, orientations and widths, with the weight  determined by the effective action.  The  explicit construction of the  domain wall network is the most recent development of the 
formalism that have been studied in the series of papers \cite{EN,EN1,NK1,NK4,NK6}, in which the domain wall defects in the homogeneous Abelian (anti-)self-dual field were taken into account either implicitly or 
in an  explicit but simplified form with the spherical domains.  The practical calculations in the next sections will be done within combined implementation of domain model given in paper~\cite{NK4}: propagators in the quark loops are taken in the approximation of the homogeneous background field and the quark loops are averaged over    the background field,  the correlators of the  background field  are calculated in the spherical domain approximation.

\begin{figure}\centering
\sidecaption
\includegraphics[scale=.9]{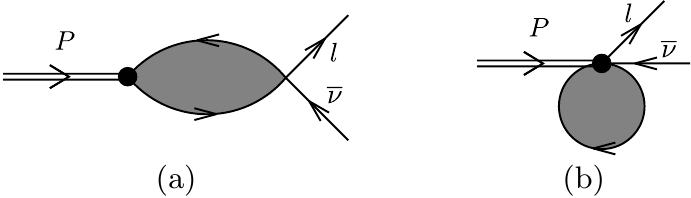}
\hspace*{5mm}\caption{Diagrams contributing to leptonic decay constants $f_P$. \label{weak_decay_diagrams}}
\end{figure}

\begin{figure}\centering
\sidecaption
\includegraphics[scale=.9]{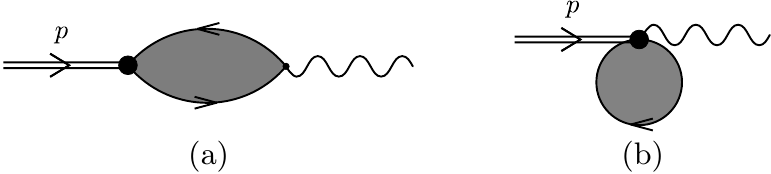}
\hspace*{5mm}\caption{Diagrams contributing to $V\to\gamma$ transition constants 
$g_{V\gamma}$.\label{g_rho_gamma_diagrams}}
\end{figure}

\section{Meson properties}
\label{section2}

 The truncated QCD functional integral can be rewritten in terms of the   composite colorless meson fields $\phi_{\cal Q}$ by means of  the standard bosonization procedure: introduce the auxiliary meson fields, integrate out the quark fields, perform the orthogonal transformation of the auxiliary fields that diagonalizes the quadratic part of the action and, finally, rescale the  meson fields to provide the correct residue of the meson propagator at the pole corresponding to its physical mass (if any).  
More details can be found in Ref.~\cite{EN,EN1,NK4}. 
The result can be written in the following compact form~\cite{NV1}
\begin{eqnarray}
&&Z={\cal N}
\int D\phi_{\cal Q}
\exp\left\{-\frac{\Lambda^2}{2}\frac{h^2_{\cal Q}}{g^2 C^2_{ \mathcal Q}}\int d^4x 
\phi^2_{\cal Q}(x)
-\sum\limits_{k=2}^\infty\frac{1}{k}W_k[\phi]\right\},
\label{meson_pf}\\
&&W_k[\phi]=
\sum\limits_{{\cal Q}_1\dots{\cal Q}_k}h_{{\cal Q}_1}\dots h_{{\cal Q}_k}
\int d^4x_1\dots\int d^4x_k
\Phi_{{\cal Q}_1}(x_1)\dots \Phi_{{\cal Q}_k}(x_k)
\Gamma^{(k)}_{{\cal Q}_1\dots{\cal Q}_k}(x_1,\dots,x_k),
\label{effective_meson_action}
\\
 &&\Phi_{{\cal Q}}(x)=\int \frac{d^4p}{(2\pi)^4}e^{ipx}{\mathcal O}_{{\mathcal Q}{\mathcal Q}'}(p)\tilde\phi_{{\mathcal Q}'}(p), \  .
 \label{Pphi}
\end{eqnarray}
where condensed index $\mathcal{Q}$ denotes all relevant meson quantum numbers and indices. Integration variables $\phi_{\mathcal Q}$ in the functional integral \eqref{meson_pf} correspond to the physical meson fields that diagonalize the quadratic part of the effective meson action \eqref{effective_meson_action} in momentum representation, which is achieved by means of orthogonal transformation ${\mathcal O}(p)$.

\begin{figure}
\begin{centering}
\sidecaption
\includegraphics[width=0.3\textwidth]{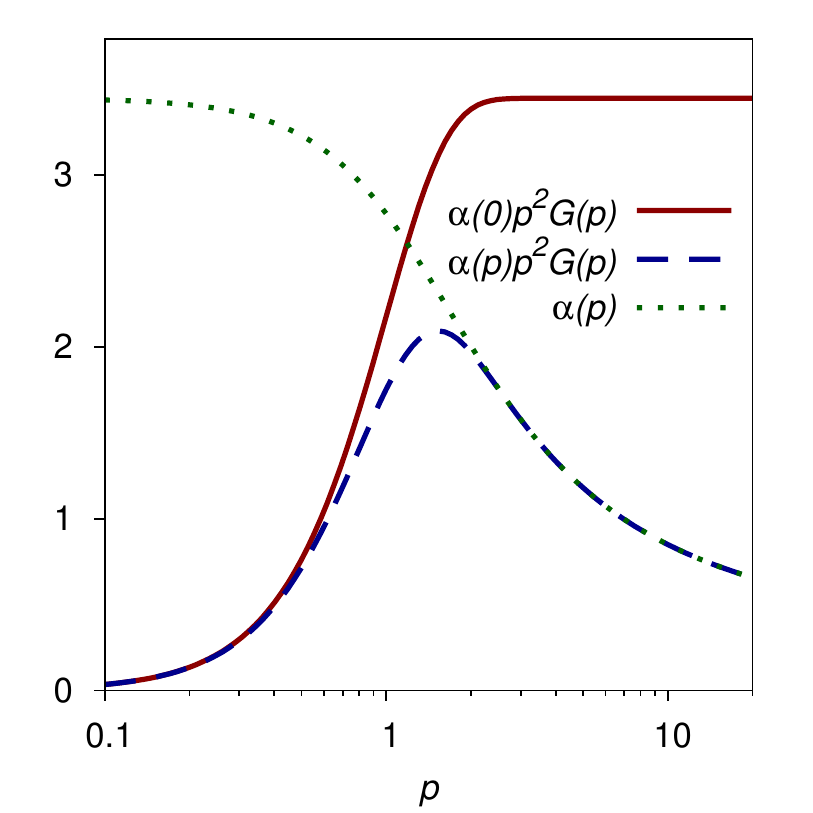}
\caption{Momentum dependence of the gluon  dressing function  without (solid line)  and with (dashed line) accounting for the running of the strong coupling constant $\alpha_s(p)$  (dotted line). The dashed line qualitatively reproduces the shape of the Landau gauge dressing function of gluons calculated within the functional renormalization group~\cite{Jan2015} and  Lattice QCD~\cite{Bowman, Ilgenfritz} as well as a part of the input gluon propagator used in the approaches based on combined  Dyson-Schwinger and Bethe-Salpeter equations~\cite{Fischer:2014xha,Dorkin:2014lxa}.  \label{gluon_fig}}
\end{centering}
\end{figure}

Interactions between physical meson fields  $\phi_{\mathcal Q}$ are  described  by  $k$-point nonlocal vertices $\Gamma^{(k)}_{\mathcal{Q}_1\dots \mathcal{Q}_2}$. subsequently tuned to the physical meson representation by means of  corresponding orthogonal transformations ${\mathcal O}(p)$. 
As is illustrated in Fig.~\ref{diagrams},  vertices $\Gamma^{(k)}$ are expressed \textit{via}  1-loop  diagrams  $G^{(k)}_{{\cal Q}_1\dots{\cal Q}_k}$ which include  nonlocal quark-meson vertices  and quark propagators (for their explicit analytical form see~\cite{NV1}).

The mass spectrum $M_\mathcal{Q}$  of mesons  and quark-meson coupling constants  $h_{\cal Q}$ are determined by the quadratic part of the effective meson action \textit{via} equations
\begin{equation*}
1=
\frac{g^2}{\Lambda^2}C^2_{\cal Q}\tilde \Pi_{\cal Q}(-M^2_{\cal Q}|B),
\  \
h^{-2}_{\cal Q}=
\frac{d}{dp^2}\tilde\Pi_{\cal Q}(p^2)|_{p^2=-M^2_{\cal Q}}, 
\end{equation*}
where $\tilde\Pi_{\cal Q}(p^2)$ is the diagonalized two-point correlator $\tilde\Gamma^{(2)}_{\cal QQ'}(p)$  put on mass shell.
Above  definition  of the meson-quark coupling constant $h_{\cal Q}$ provides correct residue at the pole. 

The results of calculations are shown in Fig.~\ref{fig-masses} and Table~\ref{constants}.
An overall accuracy of description is 10-15 percent in the lowest order calculation
 achieved with the minimal for QCD  set of parameters: infrared limits of renormalized strong coupling constant $\alpha_s$ and quark masses $m_f$, scalar gluon condensate $\langle g^2F^2\rangle$ as a fundamental scale of QCD
 and topological susceptibility of pure QCD without quarks.  This last parameter can be related to the mean size $R$ of  domains. 
 
\section{Discussion}
\label{section3}

It is interesting to take a look at the properties of gluon propagator in the present approach  in view of the known functional form of quark  and Landau gauge gluon propagators calculated within the functional renormalization group, Lattice QCD and Dyson-Schwinger equations \cite{Bowman,Ilgenfritz,Jan2015,Fischer:2014xha,Dorkin:2014lxa}.  
As it can be seen from  Fig.~\ref{gluon_fig} the shape of the gluon dressing function in the background field under consideration is in qualitative agreement with the known results of functional RG, DSE and lattice QCD. 
For more detailed comparison  we refer to the paper~\cite{NV1}.

For conclusion we would like to outline few  problems to be addressed within the domain model of QCD vacuum. The most important conceptual question relates to identification of a  mechanism for stabilization of a  finite mean size of domains. Our preliminary estimates  indicated that the competition between gluon and ghost contributions and the  contribution of the quark lowest eigenvalues  to the free energy density of the finite domain  could lead to the appropriate stabilization. This issue has to be studied carefully.    Another important for phenomenological applications problem is the  accurate description of the $n-$pont correlators in the random domain wall network ensemble which can be achieved by numerical methods. Random spherical domain ensemble is a rather rough approximation, and it has to be improved. 
Domain model of QCD vacuum offers an appealing way to study the deconfinement transition in terms of the explicit degrees of freedom which are active or suppressed in different regimes (high energy density, high baryon charge density, strong electromagnetic fields). A preliminary consideration of the relevant features of the model can be found in \cite{NV}.

%
%
%

\end{document}